\PassOptionsToPackage{dvipsnames}{xcolor}
\documentclass[
   aps,prl,twocolumn,
   reprint,
   superscriptaddress,
   nofootinbib,
   longbibliography,
   floatfix,
   amssymb
]{revtex4-1}

\usepackage{standalone}

\usepackage{amsmath}
\usepackage{amsfonts}
\usepackage{graphicx}
\usepackage{bm}
\usepackage{nicefrac} 
\usepackage{mathtools} 
\usepackage[normalem]{ulem}
\usepackage{tikz}
\usetikzlibrary{
    shapes.geometric
}

\usepackage[hypcap=false]{caption}
\usepackage[dvipsnames]{xcolor}
\definecolor{linkColor}{rgb}{0,0.3,0.7}
\usepackage{hyperref}
\hypersetup{colorlinks=true,
           allcolors=linkColor,
           pdfborder={0 0 0},
           pdfencoding=auto}

\usepackage{ragged2e} 

\newcommand{\eqPuncDistance}{\ }
\newcommand{\materialsAndMethodsLabel}{\emph{Materials and Methods}}

\begin{document}

\title{Delay-facilitated self-assembly in compartmentalized systems}

\author{Severin Angerpointner}
\thanks{S.A. and R.S. contributed equally.}
\author{Richard Swiderski}
\thanks{S.A. and R.S. contributed equally.}
\affiliation{Arnold Sommerfeld Center for Theoretical Physics and Center for NanoScience, Department of Physics, Ludwig-Maximilians-Universit\"at M\"unchen, Theresienstra\ss e 37, D-80333 Munich, Germany}

\author{Erwin Frey}
\email[Corresponding author: ]{frey@lmu.de}
\affiliation{Arnold Sommerfeld Center for Theoretical Physics and Center for NanoScience, Department of Physics, Ludwig-Maximilians-Universit\"at M\"unchen, Theresienstra\ss e 37, D-80333 Munich, Germany}
\affiliation{Max Planck School Matter to Life, Hofgartenstraße 8, 80539 Munich, Germany}

\begin{abstract}
Self-assembly processes in biological and synthetic biomolecular systems are often governed by the spatial separation of biochemical processes.
While previous work has focused on optimizing self-assembly through fine-tuned reaction parameters or using phase-separated liquid compartments with fast particle exchange, the role of slow inter-compartmental exchange remains poorly understood.
Here, we demonstrate that slow particle exchange between reaction domains can enhance self-assembly efficiency through a cooperative mechanism: \textit{delay-facilitated assembly}.
Using a minimal model of irreversible self-assembly in two compartments with distinct reaction and exchange dynamics, we identify scenarios that maximize yield and minimize assembly time, even under conditions where isolated compartments would fail to facilitate any self-assembly.
The mechanism relies on a separation of timescales between intra-compartmental reactions and inter-compartmental exchange and is robust across a wide range of geometries, including spatially extended domains with diffusive transport.
We demonstrate that this effect enables geometric control of self-assembly processes through compartment volumes and exchange rates, eliminating the need for fine-tuning local reaction rates.
These results offer a conceptual framework for leveraging spatial separation in synthetic self-assembly design and suggest that biological systems may use slow particle exchange to improve assembly efficiency.
\\

\textit{A video summary of this paper can be found at \url{https://doi.org/10.5446/72056}.}
\end{abstract}

\maketitle

Self-assembly describes how molecular subunits form functional macroscopic structures, driven by specific molecular interactions and environmental conditions.
These processes are crucial for living cells, e.g., the assembly of virus capsids~\citep{Zlotnick2011VirusCapsids, Perlmutter2015VirusCapsids}, ribosomes~\citep{Basler2018RibosomeAssemblyEukaryotes, Shajani2011RibosomeBacterial}, the flagella apparatus~\citep{ChevanceFlagellumOverview, LiFlagellumEColi}, or bacterial micro compartments~\citep{Kerfeld2018BMCreview}.
Understanding the fundamental principles of self-assembly also guides recent developments in synthetic biology and nanotechnology, for example in the design of nanoscale containers for virus trapping~\citep{siglProgrammableIcosahedralShell2021, monferrerDNAOrigamiTraps2023} or molecule delivery~\citep{khmelinskaia_structure-based_2021, jiangDNAOrigamiMolecular2024}.
One of these principles is the separation of timescales between the formation of stable nuclei, i.e., thermodynamically (meta-)stable intermediate structures, and subsequent growth of structures, to avoid kinetic traps~\citep{zlotnickTheoreticalModelSuccessfully1999, morozov_assembly_2009, hagan_understanding_2010, haganMechanismsKineticTrapping2011, keThreeDimensionalStructuresSelfAssembled2012, weiComplexShapesSelfassembled2012, reinhardtNumericalEvidenceNucleated2014, gartnerStochasticYieldCatastrophes2020}.
Hence, time-efficient self-assembly often requires fine tuning of particle densities or binding energies, or sophisticated design strategies~\citep{jacobsRationalDesignSelfassembly2015, rovigattiSimpleSolutionProblem2022, pintoDesignStrategiesSelfassembly2023, evansDesigning3DMulticomponent2024}.

Compartmentalization is another general principle that living systems rely on across different scales and for different functions~\citep{phillipsPhysicalBiologyCell2013compartments, bananiBiomolecularCondensatesOrganizers2017, greeningFormationFunctionBacterial2020}.
At the cellular level, compartments can locally accelerate or inhibit biochemical processes which play an important role in the assembly of multi-protein complexes.
Examples are the separation of the nucleolus into multiple liquid phases for the assembly of ribosomal subunits~\citep{LafontaineNucleolusOverview, quinodoz2024mRNAspatiallySeparated}, the co-localization and co-translational assembly in translational factories~\citep{Bernardini2023CotranscriptionTFIID, Crawford2024CotranslationReview}, and the interface of an enzymatic core in the ``cargo first'' assembly of some bacterial microcompartments~\citep{Kerfeld2018BMCreview, juodeikis2020encapsulatingPeptidesBMCAssembly, Yang2022PdUAssemblyPathway, abeysinghe2024InterfacialBMCAssembly, MohajeraniHagan2018MDShellAssembly}.
This high degree of spatial organization in nature has also inspired efforts to utilize compartmentalization in bio-engineering---for example, by designing \textit{de novo} micro compartments~\citep{Planamente2019BioengineeringBMCReview}, designing assembly pathways that utilize compartments \textit{in vivo}~\citep{Yang2021AssemblyInLysosomes, Wang2024OrganelleMediatedAssembly}, or by creating functional surfaces like DNA brushes that localize gene expression in microfluidics~\citep{karzbrun_programmable_2014, vonshakProgrammingMultiproteinAssembly2020}.

These examples highlight the importance of spatial separation into compartments for self-assembly processes and raise the following key questions:
Under which circumstances does localized acceleration or inhibition of reactions benefit self-assembly processes? 
Which (geometric) parameters control the assembly and what are the underlying effects governing its dynamics?

Under the assumption of fast particle exchange, it was shown that the size and composition of liquid condensates can be tuned to improve yield and assembly time of structures that either co-localize with the liquid condensate~\citep{haganSelfassemblyCoupledLiquidliquid2023, laha_chemical_2024} or phase separate on their own~\citep{bartolucci_interplay_2024}.
While the assumption of fast particle exchange is well-

\noindent{}justified for liquid condensates inside of cells~\citep{haganSelfassemblyCoupledLiquidliquid2023}, the exchange dynamics between biological or synthetic compartments can also happen on similar or even slower timescales compared to the assembly reactions.
Examples include microfluidic chambers connected by narrow channels~\citep{karzbrun_programmable_2014, tayar_synchrony_2017} and exchange through membranes or protein shells regulated by pores~\citep{Chowdhury2015SelectiveProteinShells, Lee2017ShellDiffusionMutation, chowdhury_engineering_2019, tasneem_how_2022}. Furthermore, molecular crowding and excluded volume effects can lower the mobility inside dense condensates~\citep{dix_crowding_2008, schmit_physical_2024, frechetteComputerSimulationsShow2025}. 
The resulting assembly rates can thus be limited by both the intra-compartment mobility and the diffusive exchange with the low-density bulk.

In this study, we focus on the dynamic effects of slow particle exchange on self-assembly processes in spatially extended systems.
In a conceptual two-compartment model we identify a cooperative effect of spatial separation, which we term \emph{delay-facilitated assembly}:
An optimal, finite particle exchange rate can maximize yield and minimize the assembly time, even in conditions that would not allow for effective assembly in isolated compartments.
The mechanism is remarkably generic across different geometries, reaction kinetics, and exchange dynamics, which we show in simulations of different compartment geometries---like a bulk coupled to a catalytic surface or multiple liquid-like condensates coupled to a bulk domain.
Delay-facilitated assembly can thus provide a way of utilizing spatial separation in systems where exchange is slow compared to the assembly dynamics and concentrations stay dilute during the entire assembly process.
Analyzing this mechanism aids the design of synthetic assembly systems and adds to a broader understanding of the effects of spatial separation on self-assembling systems in nature.

\begin{figure*}[ht]
    \centering
    \includegraphics[]{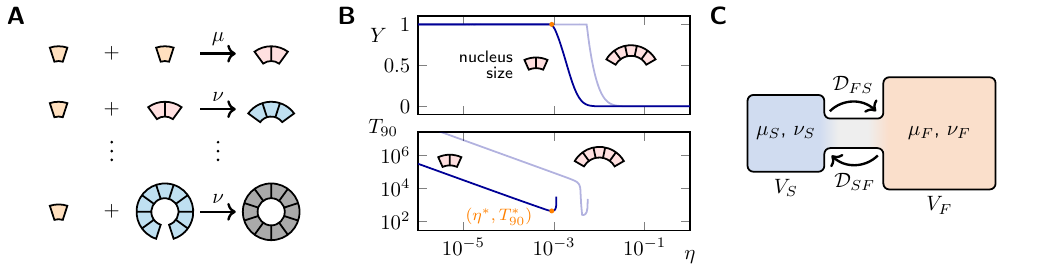}
    \caption{
    \justifying
    Model introduction and well-mixed results:
    (\textit{A}) Self-assembly model of irreversibly binding identical subunits with nuclei of size two.
    (\textit{B}) Results of well-mixed simulations for two exemplary nucleation sizes (2 and 5) when sweeping over the nucleation-to-growth ratio~$\eta$ (top: final yield~$Y$, bottom: assembly time~$T_{90}$ measured in units of the growth rate times the initial density,~$\nu\rho_0$)~\citep{gartnerStochasticYieldCatastrophes2020}.
    (\textit{C}) Sketch of the two compartment spatial setup with the biochemical and geometric control parameters.
    }
    \label{fig:model-description}
\end{figure*}

\section*{Model introduction}

To uncover general principles of self-assembly, particularly the role of spatial separation, we construct minimal models that distill essential dynamic features rather than replicate specific biological systems. These models provide a tractable and flexible framework for isolating mechanisms that are broadly shared across biological contexts, while allowing us to systematically explore how spatial organization shapes assembly processes.

\subsection*{Well-mixed assembly model}

The model we choose for most of our analysis~(Fig.~\@\ref{fig:model-description}\textit{A} and Ref.\@~\citep{gartnerStochasticYieldCatastrophes2020}), which lends ideas from aggregation models~\citep{Krapivsky_Redner_Ben-Naim_Book_2010}, builds upon various key assumptions. 
First, all subunits (e.g., individual proteins) are assumed to be identical, resulting in high-symmetry structures such as most virus capsids~\citep{reddyStructureDerivedInsightsVirus2005, zandiOriginIcosahedralSymmetry2004} and protein compartment shells~\citep{nicholsEncapsulinsMolecularBiology2017,giessenStructuralDiversityEncapsulin2024}.
Second, it is assumed that the formation of a nucleus, which can be thought of as the smallest thermodynamically favorable structure, is slow compared to subsequent structure growth---similar to classical nucleation theory~\citep{Sear2007NucleationTheory}.
Further, we assume strong subunit bindings that are irreversible on the timescale of the assembly and that structure growth stops once a final target structure is assembled.
This can be interpreted as a representation of a more general scenario where the assembly of closed structures---like a completed shell---proceeds via the formation of stable intermediate structures~\citep{gartnerDesignPrinciplesFast2024a}.
Finally, only structure growth via single subunit attachment is considered, since it has been shown in a previous analysis of similar models that higher order attachments are of minor importance~\citep{zlotnickTheoreticalModelSuccessfully1999, gartnerStochasticYieldCatastrophes2020}.

For a single well-mixed reaction compartment, the resulting mean-field dynamics for the assembly of target structures consisting of~$S$ subunits with a structure-size-independent growth rate~$\nu$ and the formation of nuclei of size two with a nucleation rate~$\mu$ are given by
\begin{subequations}\label{eq:well_mixed_equations}
    \begin{align}
        \partial_t\rho &= -2\mu\rho^2 - \nu\rho\sum_{n=2}^{S-1}\sigma_n 
        \eqPuncDistance ,
        \label{eq:well_mixed_monomers} \\
        \partial_t\sigma_2 &= \mu \rho^2 - \nu \rho \, \sigma_2 
        \eqPuncDistance ,
        \label{eq:well_mixed_dimers}\\
        \partial_t\sigma_{2<n<S} &= \nu \rho \, (\sigma_{n-1}-\sigma_n)
        \eqPuncDistance ,
        \label{eq:well_mixed_polymers}\\
        \partial_t\sigma_S &= \nu \rho \, \sigma_{S-1} \eqPuncDistance .
        \label{eq:well_mixed_targets}
    \end{align}
\end{subequations}
Here, $\rho$ is the subunit density and $\sigma_n$ the density of structures containing~$n$ subunits, which are initialized as~${\rho(0)=\rho_0}$ and~${\sigma_n(0)=0}$.

For simplicity, the target size is kept constant at~${S=30}$.
The dependence of the assembly process on the target size has been studied extensively in~\citep{gartnerStochasticYieldCatastrophes2020, gartnerTimeComplexitySelfassembly2022} and the quantities we discuss in the following exhibit known scaling laws in $S$.
Instead we focus on the impact of the reaction rates, which we analyze by varying the \textit{nucleation-to-growth} ratio~${\eta=\mu/\nu}$.
To test the generality of our findings, an alternative assembly model of square-shaped subunits is analyzed toward the end of the main text~(Fig.\@~\ref{fig:results:highlighted-model-extensions}\textit{D}).

\subsection*{Recap of well-mixed results
\cite{gartnerStochasticYieldCatastrophes2020}}

Here and in all following scenarios, we evaluate the performance of a self-assembly process using two key observables: the \textit{final yield}~$Y$
and the \textit{assembly time}~$T_{90}$.
The yield is defined as the fraction of initial subunits bound in completed structures and the assembly time~$T_{90}$ is the time required to reach a yield of~90\%.
Importantly, the assembly time as a function of the nucleation-to-growth ratio exhibits a minimum at an ideal ratio of~$\eta^*$, with a minimal time~$T_{90}^*$, whose values depend on the nucleation size~(Fig.~\@\ref{fig:model-description}\textit{B}).

If the nucleation-to-growth ratio exceeds this ideal value, excessive nucleation leads to the formation of numerous incomplete structures, resulting in a kinetic trap where assembly is stalled due to subunit depletion~\citep{hagan_understanding_2010}.
Conversely, if nucleation is slow, it becomes the rate-limiting step and the assembly time scales as $T_{90}\propto 1/\eta$.
Achieving optimal assembly times thus demands fine-tuning of reaction rates, which in many cases requires precise control over binding energies~\citep{gartnerDesignPrinciplesFast2024a}.

\subsection*{Extension to two compartments} 

While fine-tuning reaction rates is critical for optimizing self-assembly processes, precisely controlling these rates in experimental or biological settings can be challenging or impossible.
Another way how biological systems are known to organize self-assembly processes is by creating spatially inhomogeneous environments~\citep{LafontaineNucleolusOverview, Bernardini2023CotranscriptionTFIID, Crawford2024CotranslationReview}.
This raises the question of whether fine-tuning the resulting locally varying reaction rates is necessary, or whether new, more accessible control parameters emerge.

Spatially separated acceleration or inhibition of biochemical processes is realized in a multitude of different ways \textit{in vivo} and \textit{in vitro}---opening many possibilities how self-assembly processes could be organized in space.
A list of examples include membranes accelerating reactions by increasing local densities~\citep{Kholodenko2000ConcentrationIncreaseMembrane, Leonard2023MembranesChangeReactionKinetics}, aligning reaction partners~\citep{Leonard2023MembranesChangeReactionKinetics} or by enabling conformational changes of proteins~\citep{Leonard2023MembranesChangeReactionKinetics, Elsner2019AllosteryMembrane, Ebner2017AllosteryMembrane, Levina2022AllosteryMembrane}.
Further, both \textit{in vivo} and \textit{in vitro} compartments can be utilized to localize enzymes~\citep{Agapakis2012SpatialSpearatedMetabolism, Kuechler2016CompartmentsSyntheticNature, Kerfeld2018BMCreview, smokersHowDropletsCan2024}
or to control chemical parameters like the pH value~\citep{AusserwoegerPHgradientsCondensates, kingPHGradientNucleolus, martiniereInVivoPHgradients}.

Despite this diversity of spatially inhomogeneous systems, a unifying feature is their organization into two or more biochemically distinct environments, coupled by particle exchange.
To isolate the essential consequences of this spatial organization, we begin with the simplest conceptual model that captures it:
two well-mixed compartments connected by nonspecific exchange, each with distinct reaction rates~(Fig.\@~\ref{fig:model-description}\textit{C}).
The compartments are assumed to have constant volume and biochemical properties, that is, reaction rates remain constant in time and are unaffected by the assembly process.
Further, particle exchange is unspecific and does not distinguish between structures of different sizes.

The corresponding kinetic equations extend Eq.\@~Eq.\@~\eqref{eq:well_mixed_equations} by accounting for particle exchange between the compartments.
For the subunit densities $\rho_\alpha$ ($\alpha \in \{F,S\}$), this yields
\begin{alignat}{8}
\label{eq:model:single_compartment_natural_units}
    \partial_t \rho_\alpha &= 
    && \, {-}2 \mu_\alpha \rho_\alpha^2 - \nu_\alpha \rho_\alpha \sum_{n=2}^{S-1} \sigma_{n,\alpha} 
    & &&+ \frac{\mathcal{D}_{\alpha\beta}\rho_\beta \, {-} \, \mathcal{D}_{\beta\alpha}\rho_\alpha} {V_\alpha} \, ,
\end{alignat}
where~$V_\alpha$ is the volume of compartment~$\alpha$, and~$\mathcal{D}_{\beta\alpha}$ denotes the exchange rate from compartment~$\alpha$ to~$\beta$.
The remaining kinetic equations for the structure densities~$\sigma_{n,\alpha}$ follow analogously by adding compartmental exchange to Eq.\@~\eqref{eq:well_mixed_dimers}\nobreakdash--\textup{{\normalfont (\ref{eq:well_mixed_targets})}}, such that all structures undergo compartmental exchange governed by the same rates.

These equations can be non-dimensionalized by measuring all densities with respect to the initial subunit density~${\rho_{F,0} = \rho_{S,0} = \rho_0}$ and measuring time in units of the timescale of structure growth in the fast compartment, i.e.,~${\partial_t\rightarrow \nu_F \rho_0 \, \partial_t}$.
Since asymmetric exchange simply rescales certain rates compared to symmetric exchange (\materialsAndMethodsLabel), we can additionally restrict our analysis to the symmetric case of~${\mathcal{D}_{FS}=\mathcal{D}_{SF}}$.
This results in a set of rescaled equations
\begin{subequations} \label{eq:two_compartments_rescaled}
    \begin{alignat}{8}
        \partial_t \rho_F
            &=\,
            && \phantom{\tau \bigg[} {-} 2 \eta_F \rho_F^2 
            && - \rho_F 
            && \sum_{n=2} ^ {S-1} \sigma_{n,F} \phantom{\bigg]}
            && + \frac{ \mathcal{D} ( \rho_S \,{-}\, \rho_F ) }{ \phi_F }
            \label{eq:two_compartments_rescaled_alpha} \, , \\
        \partial_t \rho_S
            &=\,
            && \tau \bigg[{-}2 \eta_S \rho_S^2
            && - \rho_S 
            && \sum_{n=2}^{S-1} \sigma_{n,S} \bigg]
            && + \frac{ \mathcal{D} ( \rho_F \, {-} \, \rho_S ) }{ \phi_S } \, ,
    \end{alignat}
\end{subequations}
which feature two distinct sets of parameters: \textit{biochemical} and \textit{geometric}.
The biochemical properties of the compartments are their nucleation-to-growth ratios~${\eta_\alpha = \mu_\alpha/\nu_\alpha}$ and the relative growth rate~${\tau=\nu_S/\nu_F\leq1}$, which describes how much slower structures grow in the slow~($S$) compared to the fast~($F$) compartment.
The geometric parameters describe the structural properties of the system.
These include the total volume~${V=V_F+V_S}$, the volume fractions~${\phi_\alpha = V_\alpha/V}$ and the dimensionless exchange parameter~${\mathcal{D} = \mathcal{D}_{FS}/(\nu_F\rho_0 V)}$.
The particle exchange terms are weighted with the respective volume fractions to ensure mass-conserving particle exchange and reflect that densities in the smaller compartment are more strongly affected by particle exchange than the densities in the larger compartment.
The equations for the structure densities~$\sigma_{n,\alpha}$ follow analogously, with added exchange terms reflecting the same inter-compartmental coupling.

In a biological or experimental system, the geometric parameters can depend on various physical factors. 
For instance, volume fractions may be influenced by condensate sizes or surface-to-bulk ratio, while the exchange rate could be determined by diffusion speed, membrane attachment kinetics, or physical constraints such as the diameter, length or permeability of pores. 
\begin{figure*}[h!]
    \centering
    \includegraphics[]{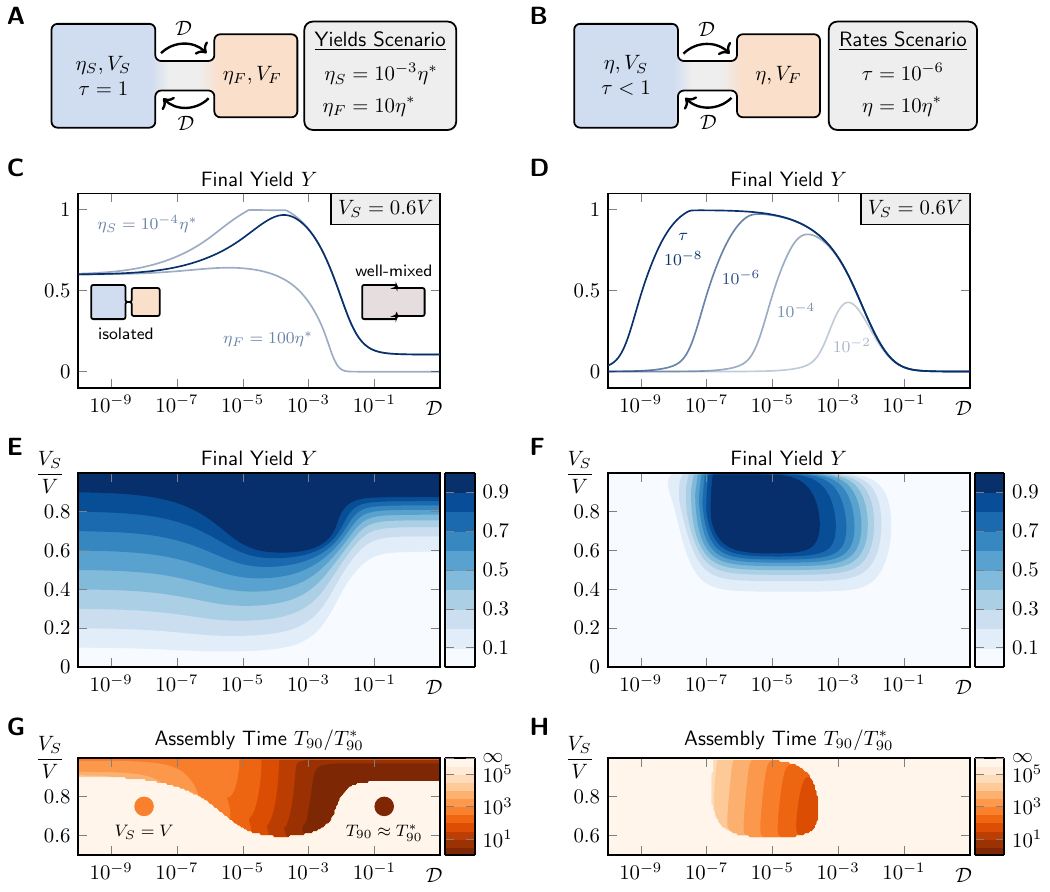}
    \caption{
        \justifying
        Main results for the yields and rates scenario in two compartments:
        (\textit{A}, \textit{B}) Non-dimensionalized control parameters and exemplary parameter choices for the remaining panels for the yields scenario and rates scenario, respectively.
        (\textit{C}) Final yield as function of compartment exchange rates~$\mathcal{D}$ (solid line) and with altered nucleation-to-growth ratios (shaded lines) at ${V_S = 0.6V}$ in the yields scenario.
        (\textit{D}) Final yield as function of compartment exchange rates~$\mathcal{D}$ for different growth ratios $\tau$ at ${V_S = 0.6V}$ in the rates scenario.
        (\textit{E}) Contour plot of the final yield for varying exchange rates~$\mathcal{D}$ and volume ratios $V_S/V$ in the yields scenario.
        (\textit{F}) Same for rates scenario.
        (\textit{G}) Contour plot for the assembly time in the yields scenario normalized with the ideal well mixed assembly time $T_{90}^*$.
        The two inset points are a guide for the eye to compare with the times reached with only the slow (high-yield) compartment~(${V_S=V}$) and the ideal well-mixed time which is realized if ${\bar\eta=\eta^*}$ and ${\mathcal{D}=\infty}$.
        (\textit{H}) Same for rates scenario.
        (\textit{E}--\textit{H}) The data for ${V_S=0}$ (${V_S=V}$) corresponds to a single well-mixed system with the parameters of the fast (slow) compartment.
        }
    \label{fig:results:main-results}
\end{figure*}
\subsection*{Compartment geometry and self-assembly scenarios} 
As fine-tuning reaction rates is often unfeasible and our goal is to understand the impact of the geometric parameters, we consider scenarios with two distinct biochemical environments where only the geometric parameters vary, while the biochemical parameters $\eta_F$, $\eta_S$, and $\tau$ remain fixed. 
This prompts two key questions: Under which biochemical conditions can geometric control enhance assembly yield or speed?
And what are the corresponding optimal geometric parameter ranges?
Since a single compartment optimized for fast and robust assembly would render the second redundant, we assume suboptimal biochemical conditions in each compartment:  
Either assembly is slow (${\eta_\alpha \ll \eta^*}$) or ineffective (${\eta_\alpha \gg \eta^*}$) in isolation, where~$\eta^*$ is again the optimal nucleation-to-growth ratio of a single well-mixed compartment.

Two distinct scenarios arise, distinguished by differences in yield or reaction rates in the two compartments. 
In the \textit{yields scenario}~(Fig.\@~\ref{fig:results:main-results}\textit{A}), both compartments share the same overall timescale (${\tau = 1}$). 
One compartment supports effective structure formation (high-yield) but suffers from slow nucleation~(${\eta_S \ll \eta^*}$), resulting in long assembly times. 
The other (low-yield) compartment exhibits faster nucleation~(${\eta_F > \eta^*}$) but does not support effective self-assembly on its own.
In the second scenario, the rates scenario~(Fig.\@~\ref{fig:results:main-results}\textit{B}), both compartments have reaction conditions unfavorable for assembly (low-yield), with identical nucleation-to-growth ratios~(${\eta_F = \eta_S = \eta > \eta^*}$). However, one compartment features a significantly slower reaction timescale~(${\tau \ll 1}$).
As we discuss in the following, these cases illustrate how geometric control can improve self-assembly performance despite suboptimal reaction rates.

\begin{figure*}[ht!]
    \centering
    \includegraphics[]{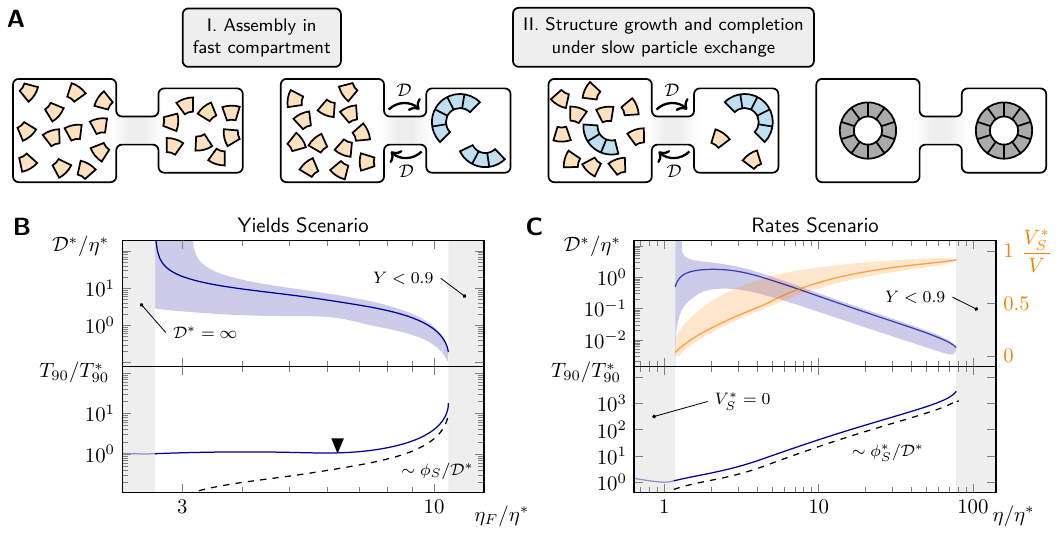}
    \caption{
    \justifying
    Optimization of assembly time:
    (\textit{A}) Sketch of the dynamics for delay-facilitated assembly: Starting from homogeneous initial conditions (leftmost sketch), assembly in the fast compartment is dominant and its subunits deplete rapidly~(\textrm{I}), resulting in unfinished structures in the fast compartment and subunits in the slow compartment (second from left).
    Subsequently, the assembly continues via particle exchange and structure growth under supply of subunits~(\textrm{II}), resulting in finished structures
    (\textit{B}). Optimal exchange parameter $\mathcal{D}^{*}$ (top) and corresponding minimal assembly time $T_{90}$ (bottom) for a system in the yields scenario with fixed ${\eta_S=10^{-3}\eta^*}$ and $V_S=0.6V$ (same as in Fig.\@~\ref{fig:results:main-results}\textit{A}) and varying ${\eta_F}$.
    The shaded region around the $\mathcal{D}^*$ curve shows the range of parameters where the reached assembly time is at most twice the optimal time.
    The regions where the optimal solution is a well-mixed compartment ($\mathcal{D}^* = \infty$, left region) and where the yield never reaches $90\%$ (right region) are shaded in gray.
    The black triangle (\protect\tikz{\protect\node[isosceles triangle, rotate=-90, scale=0.5, fill=black] {};}) marks the estimate for $\eta_F$ where the fast compartment nucleates $90\%$ of the required target structures in the initial assembly step (\textrm{I}).
    (\textit{C}) Optimal exchange parameter $\mathcal{D}^{*}$ (top, left axis) and volume fraction $V_S^*/V$ (top, right axis) and corresponding minimal assembly time $T_{90}$ (bottom) for a system in the rates scenario with fixed ${\tau=10^{-6}}$ (same as in Fig.\@~\ref{fig:results:main-results}\textit{B}) and varying ${\eta}$.
    The shaded regions around the $\mathcal{D}^*$ and $V_S^*$ curves show the range of parameters where the reached assembly time is at most twice the optimal time (keeping $V_S^*$ and $\mathcal{D}^*$ at the optimal values, respectively).
    The regions where the optimal solution is a well-mixed compartment ($\mathcal{D}^* = \infty$, left region) and where the yield never reaches $90\%$ (right region) are shaded in gray.
    }
    \label{fig:results:optimizing-control-parameters}
\end{figure*}

\section*{Two-compartment Results}

\subsection*{Intermediate exchange parameter optimizes final yield}
In the following the biochemical parameters are held fixed---corresponding either to the yields scenario~(Fig.\@~\ref{fig:results:main-results}\textit{A}) or to the rates scenario~(Fig.\@~\ref{fig:results:main-results}\textit{B})---and the self-assembly is controlled exclusively through variations of geometric parameters.
Strikingly, in both scenarios the final yield is not a monotonic function of the exchange parameter, but exhibits a maximum for intermediate values~(Figs.\@~\ref{fig:results:main-results}\textit{C}\nobreakdash--\textit{D}).
It is particularly remarkable that even in the rates scenario, where in isolation structure formation is suppressed in both compartments, almost perfect yield can be achieved if the reaction timescales are sufficiently different.
Ergo, coupling two compartments can lead to a cooperative effect which improves the efficiency of the self-assembly process beyond an interpolation between the two limiting cases of zero and infinite exchange.
This cooperative effect is present at all volume fractions, but strongest if the slow compartment (slow nucleation rate) is the larger one, i.e.,~${\phi_S > 0.5}$, which leads to an extended range of geometric parameters with high final yield~(Figs.\@~\ref{fig:results:main-results}\textit{E}\nobreakdash--\textit{F}).

Analyzing the corresponding assembly times shows that in the yields scenario~(Fig.\@~\ref{fig:results:main-results}\textit{G}) for a certain range of volume fractions, fast exchange~(darkest region for~$\mathcal{D} > 10^{-1}$) leads to assembly times close to the optimal well-mixed assembly time~$T_{90}^*$.
Importantly, comparable assembly times are also possible if particle exchange is much slower compared to growth (darkest region for~${10^{-3} < \mathcal{D} < 10^{-1}}$).
Even though this is not possible in the rates scenario~(Fig.\@~\ref{fig:results:main-results}\textit{H}) where assembly times are always significantly larger than~$T_{90}^*$, both scenarios exhibit a parameter regime where particle exchange is rate-limiting and assembly times increase as~${T_{90}\propto1/\mathcal{D}}$~(Figs.\@~\ref{fig:results:main-results}\textit{G}\nobreakdash--\textit{H} and~\ref{si:fig:example_times_with_scaling_both_scenarios}).

\subsection*{\textit{Delay-facilitated assembly} mechanism}

To interpret these results we characterize the system using three key timescales:
inter-compartment exchange~(${\tau_\mathrm{ex}\propto 1/\mathcal{D}}$), subunit depletion in the slow compartment~($\tau_S$) and subunit depletion in the fast compartment~($\tau_F$), where $\tau_S$ and $\tau_F$ are determined by the reaction kinetics in isolated compartments.
A more detailed discussion and some analytical estimates for these timescales is provided in the \materialsAndMethodsLabel{} section.

Two important limiting cases can be identified: quasi-instantaneous~(${\tau_\mathrm{ex} \ll \tau_F}$) and negligible~(${\tau_\mathrm{ex} \gg \tau_S}$) particle exchange.
In the first case, differences in compartment composition are quickly equalized and the two compartments constitute a single, effectively well-mixed compartment with volume-averaged nucleation and growth rates~$\bar\mu$ and~$\bar\nu$, leading to an averaged nucleation-to-growth rate of
\begin{equation} \label{eq:average_eta}
    \bar\eta = \frac{\bar\mu}{\bar\nu} = \frac{\mu_F V_F + \mu_S V_S}{\nu_F V_F + \nu_S V_S} \eqPuncDistance .
\end{equation}
In the yields scenario this results in an averaged nucleation-to-growth ratio of~$\bar\eta\approx\phi_F\eta_F$; thus, the final yield for fast particle exchange increases monotonically with the volume~$\phi_S$ of the slow (high-yield) compartment.
Since both compartments feature the same nucleation-to-growth ratio~${\eta > \eta^*}$ in the rates scenario, the averaged nucleation-to-growth ratio~${\bar\eta = \eta}$ is too large and the final yield for fast particle exchange is always zero.
In the opposite case of negligible particle exchange, the compartments decouple and the final yield is equal to the volume fraction of the slow (high-yield) compartment in the yields scenario~(Fig.\@~\ref{fig:results:main-results}\textit{E}) and zero in the rates scenario~(Fig.\@~\ref{fig:results:main-results}\textit{F}).

In contrast, the regime where we identified a cooperative effect between the compartments and improved final yield is defined by~${\tau_F < \tau_\mathrm{ex} < \tau_S}$.
This reflects a separation of timescales where particle exchange is \textit{slower} than the reaction dynamics in the fast compartment but \textit{faster} than reactions in the slow compartment.
Starting from a homogeneous density, the dynamics are initially dominated by the reactions in the fast compartment.
This leads to an intermediate state, where the slow compartment remains unchanged and most subunits in the fast compartment are incorporated into intermediate structures.
These structures, however, remain unfinished because nucleation is too fast~(Fig.\@~\ref{fig:results:optimizing-control-parameters}\textit{A}).

Once this state is reached, particle exchange is required for further structure growth and becomes the rate-limiting step.
Since ${\tau_\mathrm{ex} < \tau_S}$, nucleation in the slow compartment is negligible and structure growth can now occur via two pathways.
In both the yields scenario and the rates scenario a subunit can move from the slow to the fast compartment and attach to an intermediate structure there.
In the case of the yields scenario, additionally an intermediate structure can move to the slow (high-yield) compartment and grow there.
In both cases, particle exchange is rate-limiting and the ensuing \textit{effective} growth rate is proportional to the speed of particle exchange, i.e.,~${\nu_\mathrm{eff}\propto\mathcal{D}}$.
On the other hand, two particles need to be transferred to the fast compartment in order to facilitate the nucleation of a new structure.
This leads to an effective nucleation rate~${\mu_\mathrm{eff}\propto\mathcal{D}^2}$ and an effective nucleation-to-growth ratio~${\eta_\mathrm{eff}\propto\mathcal{D}}$ (see \materialsAndMethodsLabel{}).
Thus, the rate of particle exchange directly affects the effective reaction kinetics: slow exchange favors the growth of intermediate structures over nucleation, thereby enhancing final yields.

In summary, we find that a separation of reaction and exchange timescales enables a cooperative assembly mechanism in which slow particle exchange suppresses nucleation while sustaining growth. 
The process involves rapid nucleation in a high-reactivity (fast) compartment, followed by subunit supply from a low-reactivity (slow) compartment that acts as a reservoir. 
Due to the critical role of slow particle exchange, we term this process \textit{delay-facilitated assembly}.

\subsection*{Effect of compartment volumes}
The efficiency of delay-facilitated assembly depends sensitively on compartment volumes.
If the fast (high-reactivity) compartment is too large, the slow compartment (reservoir) lacks sufficient subunits to complete all intermediate structures, limiting the final yield regardless of the exchange rate.
Conversely, if the fast compartment is too small, too few nuclei are formed initially, requiring additional nucleation events that slow down assembly and reintroduce nucleation as the rate-limiting step.
In the rates scenario, this effect is particularly severe:
In the limit~${\phi_{F,S} \rightarrow 0}$, either compartment becomes kinetically trapped, and the yield vanishes.

\subsection*{Assembly time optimization}
The effectiveness of delay-facilitated assembly to enhance the final yield leads to the question of how to identify the set of geometric parameters that optimize assembly times~$T_{90}$ for given biochemical environments.
While in the rates scenario this set of parameters has to lie in the regime of delay-facilitated assembly---otherwise no final yield is possible---this is not necessary in the yields scenario.
In particular, the slow (high-yield) compartment can be large enough such that the averaged nucleation-to-growth ratio (Eq.\@~\eqref{eq:average_eta}) is small enough ($\bar\eta<\eta^*$) to facilitate high yield for fast particle exchange.
In this case slow particle exchange only slows down the dynamics and identical assembly times to a well-mixed system with nucleation-to-growth ratio equal to~$\bar\eta$ are reached for instantaneous particle exchange~${\mathcal{D}\rightarrow\infty}$ (left gray region Fig.\@~\ref{fig:results:optimizing-control-parameters}\textit{B}).
This permits a simple way to optimize the assembly time for fast particle exchange by tuning the compartment volumes such that the averaged nucleation-to-growth ratio~$\bar\eta$  is equal to~$\eta^*$, the optimal nucleation-to-growth ratio in the well-mixed scenario.
This optimization always works in the yields scenario, as long as the reaction rates follow the scenario definition of a high-yield and a low-yield compartment~(${\eta_S<\eta^*<\eta_F}$).

But what happens when this optimization scheme is not possible, either because the fast (low-yield) compartment is too large in the yields scenario or due to the biochemical environment in the rates scenario?
Then, the assembly relies on the mechanism of delay-facilitated assembly and two important criteria can be identified.
First, the fast compartment's volume must not exceed the threshold where too many structures are nucleated before particle exchange becomes relevant (triangle marker in Fig.\@~\ref{fig:results:optimizing-control-parameters}\textit{B} and black line in Fig.\@~\ref{si:fig:optimal_volume_analytic_estimate_rates_scenario}).
Indeed, this threshold value defines the optimal volume~$\phi_S^*$ in the rates scenario, where the \textit{fast} nucleation in the fast compartment is optimally utilized and no new \textit{slow} nucleations in the exchange-limited part of the dynamics are required (see~\materialsAndMethodsLabel{}).
Second, the optimal exchange parameter~$\mathcal{D}^*$ is always such that during the exchange-limited phase of the delay-facilitated assembly dynamics subunits are provided to the fast compartment just slowly enough such that no superfluous nucleation events occur.
Reducing the exchange rate beyond this point linearly slows down the assembly~(${T_{90}\propto1/\mathcal{D}}$, Fig.\@~\ref{si:fig:example_times_with_scaling_both_scenarios}).
The optimization process via the exchange parameter for delay-facilitated assembly is thus conceptually and in terms of the required fine tuning identical to the optimization over the nucleation-to-growth ratio in the well-mixed scenario: 
If it is too fast, too many nuclei form, but if it is too slow, assembly times increase linearly. 
The optimal exchange parameter~$\mathcal{D}^*$ depends on the nucleation-to-growth ratio~$\eta_F$ in the fast compartment (Figs.\@~\ref{fig:results:optimizing-control-parameters}\textit{B}\nobreakdash--\textit{C}).
As~$\eta_F$ increases, particle exchange needs to become slower to counteract the increased nucleation rate.
This is, however, only sufficient if the fast compartment is not too large or until nucleation in the slow compartment becomes too fast (right gray regions in Figs.\@~\ref{fig:results:optimizing-control-parameters}\textit{B}\nobreakdash--\textit{C}, respectively).
As particle exchange from the slow to the fast compartment gets slower, it becomes the limiting process of the overall assembly and the optimal time $T_{90}$ becomes proportional to the exchange timescale~$\phi_S/\mathcal{D}$ (dashed lines in Figs.\@~\ref{fig:results:optimizing-control-parameters}\textit{B}\nobreakdash--\textit{C}).

\begin{figure*}[ht!]
    \centering
    \includegraphics[]{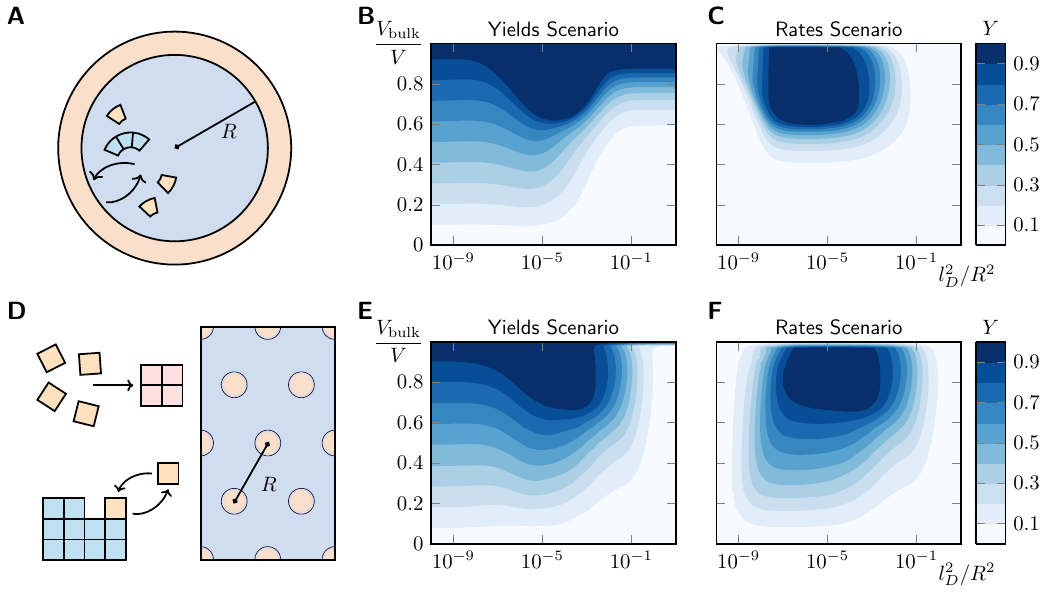}
    \caption{
    \justifying
    Delay-facilitated assembly in two spatially extended systems:
    (\textit{A}) A 2D circular domain (bulk; slow compartment) with a 1D boundary (membrane; fast compartment) where the volume fraction $V_S$ is tuned through the radius $R$ and the exchange $\mathcal{D}$ through bulk diffusion $D$ and attachment/detachment rates to/from the membrane.
    (\textit{B})~Corresponding final yield in the yields scenario (Fig.\@~\ref{fig:results:main-results}\emph{E}).
    (\textit{C}) Corresponding final yield in the rates scenario (Fig.\@~\ref{fig:results:main-results}\emph{F}).
    (\textit{D}) Different assembly system with square-shaped subunits assembling in a 2D rectangular (slow) bulk domain with (fast) circular domains, e.g., surface patterns, where the volume is set by the total area of the circular domains and exchange by diffusion and the inter-condensate distance.
    (\textit{E}) Corresponding final yield in the yields scenario (Fig.\@~\ref{fig:results:main-results}\emph{E}).
    (\textit{F}) Corresponding final yield in the rates scenario~(Fig.\@~\ref{fig:results:main-results}\emph{F}).
    The assembly times corresponding to (\textit{B}, \textit{C}, \textit{E}, \textit{F}) and are shown in Fig.\@~\ref{si:fig:assembly_times_for_model_extensions}.
    The data for ${V_{\mathrm{bulk}}=0}$ (${V_{\mathrm{bulk}}=V}$) shows the corresponding well-mixed results with the parameters of the fast (slow) domain.
    See \materialsAndMethodsLabel{} for the corresponding sets of equations and details of the numerical simulations.
    }
    \label{fig:results:highlighted-model-extensions}
\end{figure*}

\section*{Spatially extended reaction domains}

Thus far, our analysis has focused on minimal conceptual models that are simplified in both their assembly dynamics~(Fig.\@~\ref{fig:model-description}\emph{A}) and spatial structure, represented by two well-mixed compartments coupled via exchange~(Fig.\@~\ref{fig:model-description}\emph{C}).
While these simplifications were essential for elucidating the mechanism of delay-facilitated assembly, they raise an important question:
How robust is this mechanism to changes in compartment geometry, exchange dynamics, or reaction kinetics?
To address this, we consider two extensions of Eq.\@~\eqref{eq:two_compartments_rescaled} that reflect features of more realistic biological or synthetic systems.

\subsection*{Bulk-boundary geometry}

First, we investigate the effect of spatially extended compartments on the assembly model from Fig.\@~\ref{fig:model-description}\emph{A}.
The slow compartment is implemented as a 2D circular disk of radius $R$, e.g., representing the cytosol of a cell.
Its circular boundary, which could represent the cell membrane, acts as the fast compartment (Fig.\@~\ref{fig:results:highlighted-model-extensions}\emph{A}).
As before, we assume that the fast (slow) compartment is at a constant nucleation-to-growth ratio of~$\eta_F$~($\eta_S$) and that the ratio of growth rates between the slow and fast compartment is~$\tau$.
The bulk densities are now functions of space~($\rho(\bm{x})$,~$\sigma_n(\bm{x})$) and can freely diffuse.
For simplicity, we assume that all particles diffuse in the bulk with a constant coefficient~$D_n$ that decreases with the structure size~$n$, i.e.,~${D_n = D / n}$.
At the boundary, subunits and structures can attach to and detach from the boundary at constant rates, which we model as Robin boundary condition.
For simplicity, we assume the attachment and detachment rates to be the same for all structures and subunits, and to be much faster than the typical diffusion timescale;
see \materialsAndMethodsLabel{} for details of the full partial differential equation and boundary conditions.

In this system, we can identify two geometric control parameters analogous to the two-compartment scenario in Eq.\@~\eqref{eq:two_compartments_rescaled}:
The bulk radius~$R$ governs the surface-to-bulk ratio, i.e., the ratio between slow compartment and fast compartment volumes ($\phi_S$, $\phi_F$).
The dimensionless parameter
\begin{equation}
\label{eq:model-extensions:diffusive-exchange-2D}
    \mathcal{D} = \frac{l^2_D}{R^2} \eqPuncDistance ,
\end{equation}
compares the system size to the diffusive length scale and sets the effective exchange rate.
We define the diffusion length~${l^2_{D} = D / (\nu_F \rho_0)}$ by comparing the diffusivity to the timescale of structure growth with the initially homogeneous bulk density~$\rho_0$.

Importantly, exchange between bulk and boundary can also be tuned by changing the attachment and detachment rates instead of diffusion coefficients.
Similarly, instead of the volume fraction $\phi_S$, one can tune the ratio of detachment and attachment rates (see \materialsAndMethodsLabel{}).

\subsection*{2D structure assembly in liquid-like condensates}
To further test the generality of our findings, we examine a different assembly model which more closely resembles the assembly kinetics of closed virus capsids or other protein shells.
In this alternative model, square-shaped subunits form closed two-dimensional structures~(Fig.\@~\ref{fig:results:highlighted-model-extensions}\emph{D}).
In addition to nucleation and growth reactions, also the detachment of singly-bound subunits is now included in the reaction kinetics.

For incomplete shells during assembly, it is unfeasible to treat all possible configurations as states in the rate equations.
Hence, we use a set of effective rate equations derived in~\citep{gartnerDesignPrinciplesFast2024a}, which are structurally similar to Eq.\@~\eqref{eq:well_mixed_equations}, but include the effects of detachment and the particular structure morphology (see \materialsAndMethodsLabel{}).
In particular, due to subunit detachment, the first (meta-)stable nucleus now consists of four subunits~(Fig.\@~\ref{fig:results:highlighted-model-extensions}\textit{D}) instead of two~(Fig.\@~\ref{fig:model-description}\textit{A}).
Additionally, the assembly dynamics take place in a two-dimensional rectangular domain featuring a heterogeneous chemical environment, characterized by spatially varying assembly parameters  $\eta(\bm{x})$ and $\tau(\bm{x})$.
Their spatial pattern is given by equally-sized circular condensates~(Fig.\@~\ref{fig:results:highlighted-model-extensions}\emph{D}), arranged in a typical hexagonal pattern.
While we here use the term condensate, these circular domains might also represent cellular microcompartments in the cytosol or a reactive surface pattern on a 2D substrate \textit{in vitro}.
The bulk acts as the slow compartment and the circular domains act as the fast compartment(s), so either ${\eta(\bm{x} \in \text{bulk}) \ll \eta^{*}}$,~${\eta(\bm{x} \in \text{condensate}) > \eta^{*}}$, and ${\tau(\bm{x}) \equiv 1}$ (yields scenario), or ${\eta(\bm{x}) \equiv \eta > \eta^{*}}$, ${\tau(\bm{x} \in \text{bulk}) = 1}$, and ${\tau(\bm{x} \in \text{condensate}) \ll 1}$ (rates scenario).
Also, diffusion of all structures is again assumed to be constant and inversely proportional to the structure size.
For simplicity, we neglect that diffusivity typically differs between bulk and condensates~\citep{WeberReviewPhaseSeparation} and ignore effects of surface conductivity~\citep{Zhang2024ExchangeDynamicsCondensates} or protein shell permeability~\citep{tasneem_how_2022}.
See \materialsAndMethodsLabel{} for the full set of coupled partial differential equations.

We again identify two geometric control parameters.
Tuning the size of the condensates directly corresponds to tuning the fast compartments volume fraction ($\phi_F$).
The diffusion-mediated dimensionless exchange parameter takes the same form as~Eq.\@~\eqref{eq:model-extensions:diffusive-exchange-2D} and is obtained by comparing the diffusion length~$l_D$ with the distance~$R$ between the condensate centers (Fig.\@~\ref{fig:results:highlighted-model-extensions}\emph{D}).

\subsection*{Delay-facilitated assembly is generic}

In both of the spatially-extended systems described above (Fig.\@~\ref{fig:results:highlighted-model-extensions}), particle exchange is mediated through diffusion.
This implies that diffusion has to be slow, at least compared to the respective reaction timescales in the fast compartments.
Thus, we expect pronounced spatial inhomogeneities within and close to the fast compartments.
At first glance, such inhomogeneities might be expected to disrupt the mechanism of delay-facilitated assembly by breaking the assumptions of well-mixing and uniform exchange.
However, our analysis reveals otherwise: Despite the presence of spatial gradients, the essential features of delay-facilitated assembly remain intact.
Final yield (Fig.\@~\ref{fig:results:highlighted-model-extensions}) and assembly times (Fig.\@~\ref{si:fig:assembly_times_for_model_extensions}) qualitatively match the results of the conceptual two compartment system in both the yields scenario and the rates scenario (Fig.\@~\ref{fig:results:main-results}).
These findings demonstrate that the conceptual model of two coupled compartments captures the essential mechanism of delay-facilitated assembly, relevant across a wide range of spatially heterogeneous biological or synthetic systems.
The improvements in final yield and assembly times can be understood in all these systems in terms of the corresponding effective geometric control parameters: the dimensionless exchange parameter~$\mathcal{D}$ and the volume ratios~$\phi_{F,S}$ of different biochemical environments.

\section*{Discussion}

Spatial separation into compartments with slow particle exchange can improve self-assembly performance due to the effect of delay-facilitated assembly.
Coupling two compartments, each with different, suboptimal biochemical environments, can result in higher yield and faster assembly times compared to either of the individual compartments in isolation.
Controlling the exchange rate and the relative compartment volumes allows for optimizing assembly yield and times.
Remarkably, the effect is robust against variations of the geometry and the specific implementation of the exchange dynamics, e.g., through actual microfluidic compartments, bulk-boundary coupling between cytosol and cell membrane, or surface patterning.
This robustness highlights how the complexity of living systems can be partially reduced to simpler conceptual models.
In particular, using a simple two-compartment model, we identified that the mechanism behind delay-facilitated assembly is due to a separation of the reaction and exchange timescales.
Excessive nucleation in a fast compartment can be balanced by a slowly reacting reservoir that supplies subunits into the fast compartment at a steady exchange rate.

That compartments can improve assembly performance has been found in similar studies analyzing the interplay of liquid-liquid phase separation and self-assembly~\citep{haganSelfassemblyCoupledLiquidliquid2023, bartolucci_interplay_2024, frechetteComputerSimulationsShow2025}.
While the focus in these studies is on the effects of high partitioning of assembling structures into one of the compartments and fast exchange kinetics, we find a qualitatively different effect in the regime of slow exchange between the compartments.
Our results show that the benefits of compartmentalization on self-assembly are not limited to systems with fast particle exchange but extend to and can even rely on exchange-limited systems.
While we assumed here that all particles exchange between the two compartments, only the the exchange of subunits is crucial~(Fig.\@~\ref{si:fig:two_compartment_results_only_subunits_exchanging}).
The effects of size-dependent exchange on similar models of linear aggregation coupled to liquid-liquid phase separation have been studied in detail in~\citep{weberSpatialControlIrreversible2019, ponisch_aggregation_2023}.

Our conceptual model also provides clear criteria for implementing and testing delay-facilitated assembly in bio-engineering applications, by identifying and tuning effective geometric control parameters.
Examples of experimental setups are microfluidic systems designed to locally control gene expression and protein assembly~\citep{karzbrun_programmable_2014, tayar_synchrony_2017, vonshakProgrammingMultiproteinAssembly2020} where the compartment volumes are directly tunable and exchange rates depend on the distance between compartments and the properties of the solution.
Adding crowding agents could further tune diffusion rates in solution~\citep{dix_crowding_2008}.
Also, phase-separating liquid solutions are controllable to a high degree \textit{in vitro}, both regarding the rheology and chemical composition of the liquid-like domains~\citep{elbaum-garfinkleDisorderedGranuleProtein2015,bananiCompositionalControlPhaseSeparated2016}.
Functional surfaces, built with DNA origami, provide another way of spatially localizing reactions and controlling exchange through diffusion or specific binding domains on the subunits~\citep{khmelinskaia_control_2018,rosierProximityinducedCaspase9Activation2020}.
Such surfaces could even be controlled dynamically, for instance, by using electrical signals~\citep{rothfischerHighenduranceMechanicalSwitching2025}.

The mechanism of delay-facilitated assembly also provides a possible conceptual framework for studying specific self-assembly processes in nature.
For example, some metabolosomes as well as the $\beta$-carboxysome are assumed to assemble ``cargo first'' around a pre-formed enzymatic core~\citep{Kerfeld2018BMCreview, juodeikis2020encapsulatingPeptidesBMCAssembly, Yang2022PdUAssemblyPathway, abeysinghe2024InterfacialBMCAssembly, MohajeraniHagan2018MDShellAssembly}.
In this assembly process, the cytosol could be interpreted as the slow compartment (reservoir) which supplies protein shell subunits to the surface of this enzymatic core, i.e., the fast compartment.
There, the rate of shell subunit bindings is enhanced---possibly by co-localizing and aligning the convex shell subunits.
In this hypothetical scenario, the relative volume fractions and the exchange parameter could be controlled by the distance between enzymatic cores, their number and size, as well as the affinity of the shell subunits to the enzymatic core, which was numerically shown to exhibit an optimal intermediate value for shell assembly~\citep{MohajeraniHagan2018MDShellAssembly}.

The success of simple conceptual models in describing spatially separated self-assembly also suggests that other complex biological phenomena can be studied with a reductionist approach.
Such models allow the underlying effect to be understood on a generic level and notably with lower computational costs.

\section{Materials and Methods}
\subsection*{Simplified well-mixed dynamics}

For negligible density of finished structures, the assembly dynamics of a single well-mixed compartment~(Eq.\@~\eqref{eq:well_mixed_equations}) simplify to~\citep{gartnerStochasticYieldCatastrophes2020}
\begin{subequations}\label{eq:well-mixed simplified}
    \begin{align}
        \partial_t \rho &= -\eta N \rho^N - \rho \Sigma 
        \label{eq:well-mixed simplified monomers}\\
        \partial_t \Sigma &= \eta \rho^N \eqPuncDistance ,
    \end{align}
\end{subequations}
with the nucleation size~$N$, the nucleation-to-growth ratio~${\eta=\mu/\nu}$ and the total polymer density~${\Sigma = \sum_{n=N}^{S-1}\sigma_n}$.
Under the additional assumption that nucleation is slow~(${\eta \ll 1}$) one can further simplify Eq.\@~\eqref{eq:well-mixed simplified monomers} by dropping the nucleation term proportional to~$\eta$.
The solutions of the resulting set of equations with initial conditions (in rescaled units)~${\rho(0) = 1}$ and~${\Sigma(0) = 0}$ are
\begin{subequations}\label{eq:well-mixed simplified solutions}
    \begin{align}
        \rho(t) &= \cosh\left(\sqrt{\frac{N\eta}{2}} \, t\right)^{-2/N} 
        \label{eq:well-mixed simplified solution rho}\\
        \Sigma(t) &= \sqrt{\frac{N\eta}{2}} \tanh\left(\sqrt{\frac{N\eta}{2}} \, t\right) \eqPuncDistance .
    \end{align}
\end{subequations}
For small nucleation-to-growth ratios, this approximation always holds at the start of the assembly process and holds for the entire dynamics if the target size~$S$ is large enough for no structures to be completed.

Since the total number of subunits (either unbound or incorporated into bigger structures) is conserved, the average structure size $\bar l$ satisfies
\begin{equation}
    \rho(t) + \bar l(t) \Sigma(t) = 1 \eqPuncDistance .
\end{equation}
In the absorbing state all subunits are incorporated into bigger structures, yielding
\begin{equation}
    \bar l (t\rightarrow\infty)= \frac{1}{\Sigma(t\rightarrow\infty)} = \sqrt{\frac{2}{N\eta}} \eqPuncDistance .
    \label{eq:avg_length_well_mixed}
\end{equation}
In order to reach a target average length~$\bar l = L$, a nucleation-to-growth ratio proportional to~$1/L^2$ is thus needed.
This only holds for large target sizes, where the final yield can be neglected, but we assume that reaching an average polymer length~$\bar l = S'$ in an unbounded assembly (${S\rightarrow\infty}$) is equivalent to acquiring perfect yield with a finite target size~$S'$.
This leads to~${\eta^*\propto (S')^{-2}}$, which is consistent with our data~(Fig.\@~\ref{si:fig:optimal_eta_scales_as_square_of_structure_size}) and a previous more rigorous analysis of the same system~\citep{gartnerStochasticYieldCatastrophes2020}.

\subsection*{Assembly timescales in slow and fast compartment}

The dynamics of delay-facilitated assembly is governed by three different timescales:
The timescale~$\tau_\mathrm{ex}$ of particle exchange and the respective timescales~$\tau_F$ and~$\tau_S$ of monomer depletion in the fast and slow compartment.
When neglecting particle exchange, the final yield in the isolated fast compartment is always zero~(${\eta_F > \eta^*}$) and the simplified Eq.\@~\eqref{eq:well-mixed simplified} can be used.
In Fig.\@~\ref{fig:results:main-results} we further used~${\eta_F = 10\eta^* \approx10^{-2}\ll 1}$, so Eq.\@~\eqref{eq:well-mixed simplified solutions} are good approximations for the simplified assembly dynamics.
Consequently, the typical timescale for monomer depletion in the fast compartment can be inferred from Eq.\@~\eqref{eq:well-mixed simplified solution rho} as~${\tau_F \approx 1/\sqrt{\eta_F}}$.

In the rates scenario~(${\eta_S=\eta_F=\eta<\eta^*}$) the same considerations apply to the slow compartment, where time is effectively rescaled by a factor~$\tau$ leading to~${\tau_S\approx\tau/\sqrt{\eta}}$.
In the yields scenario~(${\eta_S < \eta^* < \eta_F}$), on the other hand, the final yield of an isolated slow compartment is almost 100\% invalidating the simplified model of Eq.\@~\eqref{eq:well-mixed simplified}.
In this case, nucleation is the rate limiting step and thus assembly times are inversely proportional to the nucleation rate, i.e.,~${\tau_S \propto 1/\eta_S}$.
In both scenarios,~${\tau_F < 1/\mathcal{D} < \tau_S}$ indeed approximately defines the region where particle exchange is slow enough but not too slow for the mechanism of delay-facilitated assembly to work.

\subsection*{Dynamics of delay-facilitated assembly in the rates scenario}

Under the assumptions of small nucleation-to-growth ratio, negligible amount of finished structures, and negligible reaction dynamics in the slow compartment~(${\tau \ll 1}$),~Eq.\@~\eqref{eq:two_compartments_rescaled} and the corresponding polymer dynamics simplify to
\begin{subequations}\label{eq:two-comapartments simplified}
    \begin{align}
        \partial_t \rho_F &= -\rho_F\Sigma_F + \frac{\mathcal{D} (\rho_S - \rho_F)}{\phi_F} 
        \label{eq:two-comapartments simplified monomers fast} \\
        \partial_t \Sigma_F &= \eta\rho_F^2 + \frac{\mathcal{D}(\Sigma_S - \Sigma_F)}{\phi_F} 
        \label{eq:two-comapartments simplified polymers fast} \\
        \partial_t \rho_S &= \frac{\mathcal{D} (\rho_F - \rho_S)}{\phi_S} 
        \label{eq:two-comapartments simplified monomers slow} \\
        \partial_t \Sigma_S &= \frac{\mathcal{D}(\Sigma_F - \Sigma_S)}{\phi_S} \eqPuncDistance .
        \label{eq:two-comapartments simplified polymers slow}
    \end{align} 
\end{subequations}
These are the dynamic equations of the rates scenario with nucleation size~${N = 2}$, resulting from the same approximations as~Eq.\@~\eqref{eq:well-mixed simplified}.
Here, we analyze the regime of delay-facilitated assembly as defined in the previous section~(${\tau\sqrt{\eta} \ll \mathcal{D} \ll \sqrt{\eta}}$) with initial conditions~${\rho_F(0) = \rho_S(0) = 1}$ and~${\Sigma_F(0) = \Sigma_S(0) = 0}$.

At first, the reaction kinetics in the fast compartment are much faster than particle exchange, such that~$\rho_S$ and~$\Sigma_S$ remain unchanged.
Only densities in the fast compartment change and the subunit density $\rho_F$ is depleted until the \textit{transition time}~$t_T$, where it reaches the transition value~${\rho_F(t_T) = \mathcal{D}/(\phi_F\Sigma_T)}$, with~$\Sigma_T$ being the final total polymer density of a self-assembly process in a single well-mixed compartment with nucleation-to-growth ratio~$\eta$.
After reaching $t_T$, assembly and influx from the slow compartment balance, such that the subunit density in the fast compartment remains almost constant, i.e.,~${\partial_t\rho_F=\mathcal{O}(\mathcal{D}^2)}$, and up to linear order in the exchange parameter~$\mathcal{D}$ the dynamics of the subunit density in the slow compartment~(Eq.\@~\eqref{eq:two-comapartments simplified monomers slow}) and of both polymer densities~(Eq.\@~\eqref{eq:two-comapartments simplified polymers fast}~and~Eq.\@~\eqref{eq:two-comapartments simplified polymers slow}) decouple from~$\rho_F$.
The solutions of the ensuing set of equations are given by
\begin{subequations} \label{eq:two-compartment simplified solutions}
    \begin{align}
        \rho_F(t) &= 
        \frac{\mathcal{D}\rho_S(t)}{\phi_F\Sigma_F(t)} 
        \label{eq:two-compartment simplified solution monomers fast}
        \\[1mm]
        \Sigma_F(t) &=
        \Sigma_T \left( \phi_F + \phi_S e^{-\frac{\mathcal{D} (t-t_T)}{\phi_F \phi_S}}\right)
        \label{eq:two-compartment simplified solution polymers fast}
        \\[1mm]
        \rho_S(t) &= 
        e^{-\mathcal{D} (t-t_T) / \phi_S} 
        \label{eq:two-compartment simplified solution monomers slow}
        \\[1mm]
        \Sigma_S(t) &= 
        \Sigma_T \left( \phi_F -  \phi_F e^{-\frac{\mathcal{D} (t-t_T)}{\phi_F \phi_S}}\right) \eqPuncDistance ,
        \label{eq:two-compartment simplified solution polymers slow}
    \end{align}
\end{subequations}
for~${t>t_T}$.
In Eq.\@~\eqref{eq:two-compartment simplified solution monomers fast} we used that due to the growth term~$-\rho_F\Sigma_F$, the subunit density in the fast compartment~(Eq.\@~\eqref{eq:two-comapartments simplified monomers fast}) reacts on a faster timescale proportional to~$1/\Sigma_T$ to changes in~$\rho_S$ and~$\Sigma_F$, which can be assumed to be constant on this timescale.
Consequently, it is subject to the given quasi-stationary relation.
Accordingly, the final average polymer length
\begin{equation}
    \bar l = \frac{1}{\Sigma (t\rightarrow\infty)} = \frac{1}{\phi_F \Sigma_T} \eqPuncDistance ,
\end{equation}
with the total polymer density~${\Sigma = \phi_F\Sigma_F + \phi_S \Sigma_S}$ is increased by a factor of~$1/\phi_F$ compared to a single well-mixed compartment.
Lastly, consistent with Eq.\@~\eqref{eq:well-mixed simplified} and to leading order in~$\mathcal{D}$, we define the (possibly time dependent) effective nucleation and growth rates as
\begin{subequations}
    \begin{alignat}{8}
        \mu_{\mathrm{eff}} &\coloneqq
        \frac{\partial_t\Sigma}{\rho} = 
        \frac{\eta\mathcal{D}^2}{\phi_S^2\phi_F\Sigma_T^2} 
        \left( \phi_F + \phi_S e^{-\frac{\mathcal{D} t}{\phi_F \phi_S}}\right)^{-2}
        \\[1mm]
        \nu_{\mathrm{eff}} &\coloneqq
        -\frac{\partial_t \rho + 2\partial_t\Sigma}{\rho\Sigma} = 
        \frac{\mathcal{D}}{\phi_F\phi_S\Sigma_T}
        \\[1mm]
        \eta_\mathrm{eff} &\coloneqq
        \frac{\mu_\mathrm{eff}}{\nu_\mathrm{eff}} = 
        \frac{\eta\mathcal{D}}{\phi_S \Sigma_T}  \left( \phi_F + \phi_S e^{-\frac{\mathcal{D} t}{\phi_F \phi_S}}\right)^{-2} \eqPuncDistance .
    \end{alignat}
\end{subequations}

\subsection*{Effect of asymmetric particle exchange}

In the case of asymmetric particle exchange~(${\mathcal{D}_{FS} \neq \mathcal{D}_{SF}}$ in Eq.\@~\eqref{eq:model:single_compartment_natural_units}), we restrict the initial conditions such that particle exchange is initially equilibrated:
\begin{equation}
\label{eq:methods:equilibrated_initial_conditions}
    \mathcal{D}_{SF}\rho_{F,0}
    =
    \mathcal{D}_{FS}\rho_{S,0} \eqPuncDistance .
\end{equation}

Rescaling time as for Eq.\@~\eqref{eq:two_compartments_rescaled}~(${\partial_t \rightarrow \nu_F \rho_{F,0} \partial_t}$) and measuring all densities with respect to the initial subunit density in the corresponding compartment~(${\rho_{\alpha} \rightarrow \rho_{\alpha,0}\rho_\alpha}$ and ~${\sigma_{n,\alpha} \rightarrow \rho_{\alpha,0}\sigma_{n,\alpha}}$) yields again
\begin{subequations}
    \begin{alignat}{8}
        \partial_t \rho_F
            &=
            && \phantom{\tau \bigg[} - 2 \eta_F \rho_F^2 
            && - \rho_F 
            && \sum_{n=2} ^ {S-1} \sigma_{n,F} \phantom{\bigg]}
            && + \frac{ \mathcal{D} ( \rho_S {-} \rho_F ) }{ \phi_F }
            \label{eq:two_compartments_rescaled_alpha_methods} \\
        \partial_t \rho_S
            &=
            && -\tau \bigg[2 \eta_S \rho_S^2
            && + \rho_S 
            && \sum_{n=2}^{S-1} \sigma_{n,S} \bigg]
            && + \frac{ \mathcal{D} ( \rho_F {-} \rho_S ) }{ \phi_S } .
    \end{alignat}
\end{subequations}
The only difference to Eq.\@~\eqref{eq:two_compartments_rescaled}
is that the dimensionless parameters~$\tau$,~$\mathcal{D}$ and~$\phi_\alpha$ are now given by
\begin{align}
    \tau &= \frac{\nu_S\rho_{S,0}}{\nu_F\rho_{F,0}} \eqPuncDistance , \
    \mathcal{D} = \frac{\mathcal{D}_{SF}}{N_0\nu_F} \eqPuncDistance , \
    \phi_\alpha = \frac{V_\alpha\rho_{\alpha,0}}{N_0} \eqPuncDistance ,
\end{align}
with the total number of initial subunits $N_0 = V_F \rho_{F,0} + V_S \rho_{S,0}$.
Therefore, the non-dimensionalized dynamic equations for asymmetric particle exchange are functionally identical to the ones for symmetric particle exchange, provided we start with equilibrated initial conditions satsifying~Eq.\@~\eqref{eq:methods:equilibrated_initial_conditions}.
Un-equilibrated initial conditions can be modeled by choosing the non-dimensionalized initial values for~$\rho_F$ or~$\rho_S$ different from~$1$.

\subsection*{Analytic approximations for important volume fractions}

The heuristic to find the ideal volume fractions in the regime of delay-facilitated assembly is to utilize the fast nucleation in the fast compartment as efficiently as possible: 
During the first stage of delay-facilitated assembly, where the reactions in the fast compartment dominate, exactly the amount of needed nuclei should be constructed.
In that case, no further slow nucleations in the exchange-limited regime are required.
Simultaneously not too many intermediate structures emerge, which would lead to kinetic traps.

In order for~$N_0$ initial subunits to all be incorporated into finished target structures of size~$S$ (100\% yield), ${N_\mathrm{nuc} = N_0/S}$ structures need to be nucleated.
Accordingly, in order to reach 90\% as fast as possible, the number of nucleated structures in the first stage of delay-facilitated assembly should be~${N_{\mathrm{nuc}, F} = 0.9N_0/S}$.
Using the average structure size from Eq.\@~\eqref{eq:avg_length_well_mixed}, this leads to
\begin{subequations}
    \begin{gather}
        N_{\mathrm{nuc}, F} = \frac{N_{0,F}}{\bar l_F} = \sqrt{\eta_F}N_{0,F} \overset{!}{=}\frac{0.9N_0}{S} \eqPuncDistance , \\[3mm]
        \Rightarrow \phi_F^* = \frac{0.9}{\sqrt{\eta_F}S} \eqPuncDistance,
        \label{eq:ideal_volume}
    \end{gather}
\end{subequations}
with the initial number of subunits~$N_{0,F}$ in the fast compartment.
Eq.\@~\eqref{eq:ideal_volume} approximates well the measured ideal volume fractions in the rates scenario~(Fig.\@~\ref{fig:results:optimizing-control-parameters}\textit{C} and~\ref{si:fig:optimal_volume_analytic_estimate_rates_scenario}).
It also predicts the regime of nucleation-to-growth ratios for which delay-facilitated assembly is almost equally fast as optimal well-mixed assembly in the yields scenario~(triangle marker in Fig.\@~\ref{fig:results:optimizing-control-parameters}\textit{B} at~${\eta_F=(\frac{0.9}{\phi_F S})^2}$).

Similarly, one can approximate for which values of~$\eta_F$ and~$\phi_F$ it is no longer possible to reach 90\% yield.
This is the case if the number of subunits remaining in the reservoir (slow compartment) is not sufficient to finish enough structures.
In the first stage of assembly, the intermediate structures in the fast compartment reach an average size of~${\bar l  = 1/\sqrt{\eta_F}}$; thus, on average~${S-\bar l}$ subunits are needed to finish a single structure.
Since at least~$0.9N_0/S$ structures need to be finished to reach a final yield of at least 90\%, the number of remaining subunits in the reservoir (which is approximately its initial value) needs to exceed
\begin{equation}
    N_{0,S} > \frac{0.9N_0 (S-1/\sqrt{\eta_F})}{S} \eqPuncDistance.
\end{equation}
Equivalently, for a fixed volume the final yield can only reach 90\% if
\begin{equation}
    \eta_F < S \left( 1-\frac{\phi_S}{0.9} \right) ^{-2} \eqPuncDistance . \label{eq:condition_eta_yields_scenario}
\end{equation}
Inserting the values used in~Fig.\@~\ref{fig:results:optimizing-control-parameters}\textit{B}~(${S=30}$,~${\phi_S=0.6}$ and~${\eta^*\approx8.9 \times10^{-4}}$) into Eq.\@~\eqref{eq:condition_eta_yields_scenario} yields~${\eta_F/\eta^* < 11.2}$ which fits well the 
numerically determined value of~$\eta_F/\eta^* \approx 10.7$ (start of the gray region on the right).

\subsection*{PDEs for bulk-boundary coupling}

To analyze the assembly dynamics in spatially extended systems, we choose all densities to be space-dependent and to diffuse with a structure-size-dependent diffusion constant~${D_n = D/n}$.
In the case of the circular bulk-boundary geometry (e.g., cytosol coupled to an enclosing membrane, Fig.\@~\ref{fig:results:highlighted-model-extensions}\textit{A}), both the bulk nucleation rate~$\mu$ and growth rate~$\nu$ are spatially constant.
This yields the following partial differential equations for the bulk densities~$\rho(\bm{x})$ and~$\sigma_n(\bm{x})$
\begin{subequations}\label{eq:methods:pde_bulk_boundary}
    \begin{alignat}{8}
        \partial_t \rho(\bm x ) 
        &= D \Delta \rho(\bm x ) 
        &&+ \textrm{r.h.s \eqref{eq:well_mixed_monomers}} 
        \eqPuncDistance ,
        \label{eq:methods:pde_bulk_boundary_monomers}\\
        \partial_t\sigma_n(\bm{x}) 
        &= \frac{D}{n} \Delta\sigma_n(\bm{x})
        &&+ \textrm{r.h.s \eqref{eq:well_mixed_dimers}--\textup{{\normalfont (\ref{eq:well_mixed_targets})}}}
        \eqPuncDistance . 
    \end{alignat}
\end{subequations}
In polar coordinates, we denote the radial coordinate (distance from the center of the bulk domain) by~$r$ and the polar angle by~$\varphi$.
On the boundary ($r=R$), the assembly reactions are governed by different reaction rates~$\bar\mu$ and~$\bar\nu$ and the time evolution of the boundary densities~$\bar\rho$ and~$\bar\sigma_n$ is given by the reactions and attachment to and detachment from the boundary with rates~$k_a$ and~$k_d$, respectively:
\begin{subequations}
    \begin{alignat}{8}
        \partial_t\bar\rho(\varphi) &= k_a \rho(R)
        && {-} \,  k_d\bar\rho (\varphi)
        &&+  \textrm{r.h.s \eqref{eq:well_mixed_monomers}}
        \eqPuncDistance ,
        \label{eq:methods:pde_bulk_boundary_boundary_monomers}\\
        \partial_t\bar\sigma_n(\varphi) &= k_a \sigma_n(R) \,
        &&  {-} \, k_d\bar\sigma_n (\varphi)
        &&+ \textrm{r.h.s \eqref{eq:well_mixed_dimers}--\textup{{\normalfont (\ref{eq:well_mixed_targets})}}}
         \, .
    \end{alignat}
\end{subequations}
For simplicity, we only consider the rotationally symmetric case with homogeneous intital conditions in the bulk~($\rho(\bm{x})\equiv\rho_0$) and on the boundary~($\bar\rho(\varphi) \equiv \bar\rho_0$).
Hence, we can neglect diffusion on the boundary along the~$\varphi$-direction and drop any angular dependence of densities in both domains.

At the system boundary, the bulk densities are coupled to the boundary densities via Robin boundary conditions that balance the diffusive flux with attachment to and detachment from the boundary.
In particular, the boundary conditions are given by
\begin{subequations}\label{eq:methods:pde_bulk_boundary_boundary_condition}
    \begin{align}
        D\partial_r\rho(R) &= k_d\bar\rho - k_a \rho(R) \eqPuncDistance ,
        \label{eq:methods:pde_bulk_boundary_boundary_condition_monomers} \\
        D\partial_r\sigma_n(R) &= k_d\bar\sigma_n - k_a \sigma_n(R) \eqPuncDistance .
    \end{align}
\end{subequations}
Finally, we choose the initial conditions such that bulk-membrane particle exchange is initially equilibrated, i.e., we choose~${k_a \rho_0 = k_d\bar\rho_0}$ for the initial subunit densities.
Non-dimensionalizing Eq.\@~\eqref{eq:methods:pde_bulk_boundary_monomers}, Eq.\@~\eqref{eq:methods:pde_bulk_boundary_boundary_monomers}, and Eq.\@~\eqref{eq:methods:pde_bulk_boundary_boundary_condition_monomers} by rescaling time~(${\partial_t \rightarrow \bar\nu \bar\rho_0 \partial_t}$), initial densities~(${\rho(r) \rightarrow \rho_0\rho(r/R)}$ and~${\bar\rho\rightarrow \rho_0\bar\rho}$), and space~(${\partial_r \rightarrow \partial_r/R}$) yields
\begin{subequations}
    \begin{align}
        \partial_t \rho(r) &= 
        \mathcal{D}\Delta\rho(r) - \tau \left[ \eta \rho(r)^2 - \sum_{n=1}^{S-1} \sigma_n(r) \right]
        \\
        \partial_t \bar\rho &=
         \mathcal{D}_d \left[ \rho(1) - \bar\rho \right]
        - \bar\eta \bar\rho^2 - \bar\rho \sum_{n=1}^{S-1} \bar\sigma_n
        \\
        \partial_r \rho(1) &= 
        \mathcal{D}_a \left[\bar\rho - \rho(1)\right] \eqPuncDistance , ~~ r\in[0,1] \eqPuncDistance .
    \end{align}
\end{subequations}
The corresponding dimensionless parameters are given by
\begin{subequations}
\begin{gather}
    \eta = \frac{\mu}{\nu} \eqPuncDistance , \ 
    \bar\eta = \frac{\bar\mu}{\bar\nu} \eqPuncDistance , \ 
    \tau = \frac{\nu \rho_0}{\bar\nu \bar\rho_0} \eqPuncDistance , \\
    \mathcal{D} = \frac{D}{\bar\nu \bar\rho_0R^2} \eqPuncDistance , \
    \mathcal{D}_d = \frac{k_d}{\bar\nu \bar\rho_0} \eqPuncDistance , \
    \mathcal{D}_a = \frac{k_a R}{D} \eqPuncDistance ,
\end{gather}
\end{subequations}
and the dimensionless volumes, i.e., initial subunit numbers, used in Figs.\@~\ref{fig:results:highlighted-model-extensions}\textit{B}\nobreakdash--\textit{C} are given by
\begin{subequations}
    \begin{align}
        V_\mathrm{bulk} &= \pi R^2 \rho_0 \\[1mm]
        V_\mathrm{bound} &= 2 \pi R \bar\rho_0 \\[1mm]
        \frac{V_\mathrm{bulk}}{V} &= \frac{R}{R + 2 \bar\rho_0/\rho_0} \eqPuncDistance .
    \end{align}
\end{subequations}
The effective exchange parameter is related to the diffusive length scale, $\mathcal{D}=(l_D/R)^2$, mentioned in the main text.
The non-dimensionalization of the structure densities~$\sigma_n(\bm x)$ and~$\bar\sigma$ follows accordingly, by also rescaling~${\sigma_n(r) \rightarrow \rho_0\sigma_n(r/R)}$ and~${\bar\sigma \rightarrow \bar\rho_0 \bar\sigma}$.
For the numerical results shown in Figs.\@~\ref{fig:results:highlighted-model-extensions}\textit{B}\nobreakdash--\textit{C} we used, ${\mathcal{D}_d = 10^2}$, and discretized the PDE using a finite-difference scheme over $100$ grid points along the radial dimension.
This choice of parameters implies that the volume fraction~$V_\mathrm{bulk}/V$ and the diffusive parameter~$\mathcal{D}$ can be controlled solely through changing either the system radius~$R$ or the microscopic diffusion constant~$D$.
The assembly parameters were chosen as ${\bar\eta=10\eta^*}$, ${\eta=10^{-3}\eta^*}$, ${\tau=1}$ in the yields scenario~(Fig.\@~\ref{fig:results:highlighted-model-extensions}\textit{B}) and ${\bar\eta = \eta = 10\eta^*}$ and ${\tau = 10^{-6}}$ in the rates scenario~(Fig.\@~\ref{fig:results:highlighted-model-extensions}\textit{C}).

\subsection*{PDEs for bulk-condensate coupling}

For the assembly dynamics in Fig.\@~\ref{fig:results:highlighted-model-extensions}\textit{D} we replaced the linear assembly model of Fig.\@~\ref{fig:model-description}\textit{A} with a model of identical square-shaped subunits assembling into two-dimensional structures adopted from Ref.\@~\citep{gartnerDesignPrinciplesFast2024a}.
We assume that subunits can attach to any other subunit or to a bigger structure at any available site with rate~$\nu$, singly-bound squares detach with rate~$\delta$ and doubly-bound squares are stably incorporated into the structure~(Fig.\@~\ref{fig:results:highlighted-model-extensions}\textit{C}).
This assembly is well characterized by the following effective kinetic theory~\citep{gartnerDesignPrinciplesFast2024a}:
\begin{subequations}\label{eq:floGs_equations}
    \begin{align}
        \partial_t\rho &= -N\bar\mu\rho^N - \bar\nu\rho^\gamma \sum_{n=N}^{S-1}f_n \sigma_n \\
        \partial_t\sigma_N &= \bar\mu\rho^N - \bar\nu\rho^\gamma f_N\sigma_N \\
        \partial_t\sigma_{N<n<S} &= \bar\nu\rho^\gamma(f_{n-1}\sigma_{n-1} - f_n\sigma_n) \\
        \partial_t\sigma_S &= \bar\nu\rho^\gamma f_{S-1}\sigma_{S-1}  \eqPuncDistance .
    \end{align}
\end{subequations}
Eq.\@~\eqref{eq:floGs_equations} are very similar to Eq.\@~\eqref{eq:well_mixed_equations} with an effective nucleation rate~${\bar\mu = \nu(\nu\rho_0/\delta)^{N-1}}$ and effective growth rate~${\bar\nu = \nu(\nu\rho_0/\delta})^{\gamma-1}$.
The model depends on the subunit morphology, i.e., the nucleation size~$N$, the attachment order~$\gamma$, and a structure-size-dependent combinatorial prefactor~$f_n$.
For square-shaped subunits these parameters have been found to be~$N=4$,~$\gamma=2$ and~${f_n = 5.3n}$~\citep{gartnerDesignPrinciplesFast2024a}.
This leads to the dimensionless nucleation-to-growth ratio~${\eta=(\nu\rho_0/\delta)^2}$.

For the spatially extended system we add diffusion to~Eq.\@~\eqref{eq:floGs_equations} with a structure-size-dependent diffusion constant~${D_n = D/n}$ in a rectangular geometry with no-flux boundary conditions.
We choose homogeneous initial conditions ($\rho(\bm{x}) \equiv \rho_0$) and introduce space-dependent rates~$\delta(\bm{x})$, $\nu(\bm{x})$.
Analogous to the two-compartment scenario, we define a fast and a slow environment with effective nucleation-to-growth ratios~$\eta_F$ and~$\eta_S$, respectively.
To model the condensate-like pattern in Fig.\@~\ref{fig:results:highlighted-model-extensions}\textit{D} we define this ratio as
\begin{equation}
    \eta(\bm x) = \eta_S + \sum_{i=1}^{N_d} \frac{\eta_F-\eta_S}{1+\exp[(|\bm{x}-\bm{d}_i|-r_d)/\xi]} \eqPuncDistance ,
\end{equation}
with~$N_d$ condensates, condensate positions~${\bm{d}_i}$, condensate radius~$r_d$ and interface width~$\xi$.
In the yields scenario this choice leads to constant growth rate~$\nu(\bm{x})\equiv\nu$ and space-dependent detachment rate~$\delta(\bm{x})=\nu\rho_0/\eta(\bm{x})^{1/2}$, whereas in the rates scenario it leads to both rates varying such that~$\eta(\bm{x})\equiv\eta$ stays constant, and $\nu(\bm{x})\rho_0$ varies between $1$ (fast domains) and $\tau$.
The condensate positions are given by the equidistant lattice points of a hexagonal lattice and their total volume is given by~$V_d = \pi N_d r_d^2$.
In the numeric simulations, we perform a finite-difference discretization with~${50 \times 87}$ grid points and place~$N_d=8$ condensates as shown in Fig.\@~\ref{fig:results:highlighted-model-extensions}\textit{D} with an interface width~$\xi=10^{-4}$.
The reaction parameters were chosen as $\eta_F=10^2\eta^*$ and $\eta_S=10^{-2}\eta^*$ in the yields scenario~(Fig.\@~\ref{fig:results:highlighted-model-extensions}\textit{E}), and as $\eta=100\eta^*$, $\tau = 10^{-6}$ in the rates scenario~(Fig.\@~\ref{fig:results:highlighted-model-extensions}\textit{F}).
We used structures of size~$S=30$ for which ${\eta^*=(\nu\rho_0/\delta)^2 \approx 8.5\times10^{-4}}$.

\subsection*{Numerical simulations}

Differential equations for Figs.\@~\ref{fig:model-description}\nobreakdash--\ref{fig:results:highlighted-model-extensions} were solved numerically with the Julia package DifferentialEquations.jl~\citep{rackauckas2017differentialequations}, using an implicit Runge-Kutta method (ESDIRK).
Data in Figs.\@~\ref{fig:results:highlighted-model-extensions}\textit{E}\nobreakdash--\textit{F} is partially interpolated due to sparse data at very low diffusivites.
Optimizations shown in Figs.\@~\ref{fig:results:optimizing-control-parameters},\@~\ref{si:fig:optimal_volume_analytic_estimate_rates_scenario}, and~\ref{si:fig:optimal_eta_scales_as_square_of_structure_size} were carried out using Optimization.jl~\citep{vaibhav_kumar_dixit_2023_7738525} with a combination of gradient-free and BFGS methods.
The full code to reproduce our results is available on Zenodo~\citep{code_for_publication}.

\subsection*{Acknowledgements}
\begin{acknowledgments}
    We thank Florian Gartner and Florian Raßhofer for stimulating discussions and critical review of the manuscript. 
    This work was funded by the 
    Deutsche Forschungsgemeinschaft (DFG, German Research Foundation) through the Excellence Cluster ORIGINS under Germany’s Excellence Strategy - EXC-2094 - 390783311, 
    the European Union (ERC, CellGeom, project number 101097810),
    the grant NSF PHY-2309135 to the Kavli Institute for Theoretical Physics (KITP)
    and the Chan-Zuckerberg Initiative (CZI).
\end{acknowledgments}

\bibliography{references.bib}

\renewcommand{\figurename}{Supplementary Figure}
\newcounter{offset}
\setcounter{offset}{\value{figure}}
\renewcommand{\thefigure}{S\the\numexpr\value{figure}-\value{offset}\relax}

\clearpage

\onecolumngrid

\section{Supplemental Material: \\ Delay-facilitated self-assembly in compartmentalized systems}

\begin{center}
\begin{minipage}{\textwidth}
    \centering
    \includegraphics[]{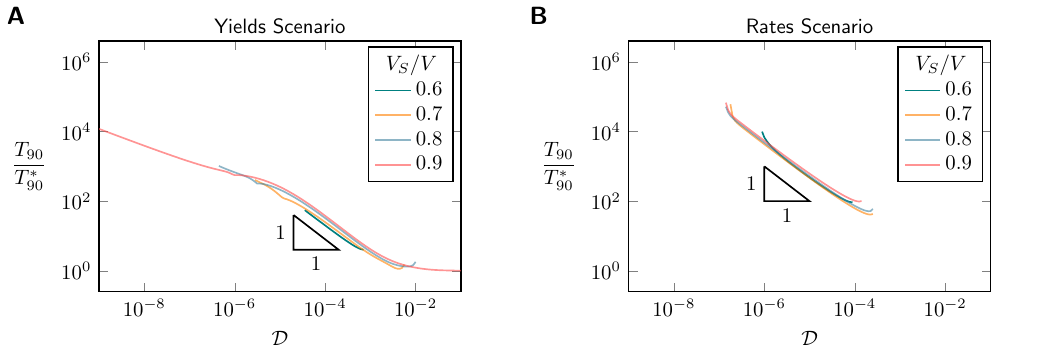}
    \captionof{figure}{
        \justifying
        Assembly times to reach a yield of 90\%~($T_{90}$) for the two-compartment model as a function of the exchange parameter $\mathcal{D}$ for selected volumes.
        (\textit{A}) Yields scenario.
        (\textit{B}) Rates scenario.
        All curves exhibit a scaling regime with ${T_{90} \sim \mathcal{D}^{-1}}$, where exchange becomes the rate-limiting process for assembly.
    }
    \label{si:fig:example_times_with_scaling_both_scenarios}
\end{minipage}
\end{center}
\vspace{3em}

\begin{center}
\begin{minipage}{\textwidth}
    \centering
    \includegraphics[]{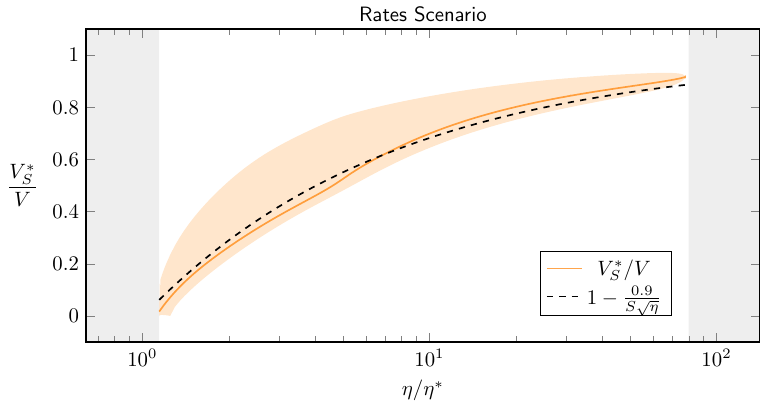}
    \captionof{figure}{
    \justifying
    Optimal volume fraction $\phi_S^*=V_S^*/V$ of a two-compartment system in the rates scenario as a function of the nucleation-to-growth ratio~$\eta$ (normalized by the optimal well-mixed nucleation-to-growth ratio~$\eta^*$), same data as Fig.\@~\ref{fig:results:optimizing-control-parameters}\textit{C}.
    The optimal values are obtained by minimizing the assembly time~$T_{90}$ over the volume fraction~$\phi_S$ and exchange rate~$\mathcal{D}$ simultaneously.
    The orange shaded region marks the range of volume fractions for each value of~$\eta$ (and corresponding optimal exchange rate) for which the assembly time~$T_{90}$ reaches at most twice the minimal value.
    The dashed line corresponds to the threshold volume mentioned in the main text,\@~$\phi_S^* = 1-\phi_F^*=1-0.9/(S\sqrt{\eta_F})$, below which too many nucleated structures form in the fast compartment (\materialsAndMethodsLabel).
    As in Fig.\@~\ref{fig:results:optimizing-control-parameters}\textit{C} the gray shaded regions mark the range where~$\eta$ is small enough for a single well-mixed compartment to outperform the two-compartment system for any exchange rate (left region), and the range where~$\eta$ is too large for the system to produce~90\% yield (right region).
    }
    \label{si:fig:optimal_volume_analytic_estimate_rates_scenario}
\end{minipage}
\end{center}

\begin{figure*}[h]
    \centering
    \includegraphics[]{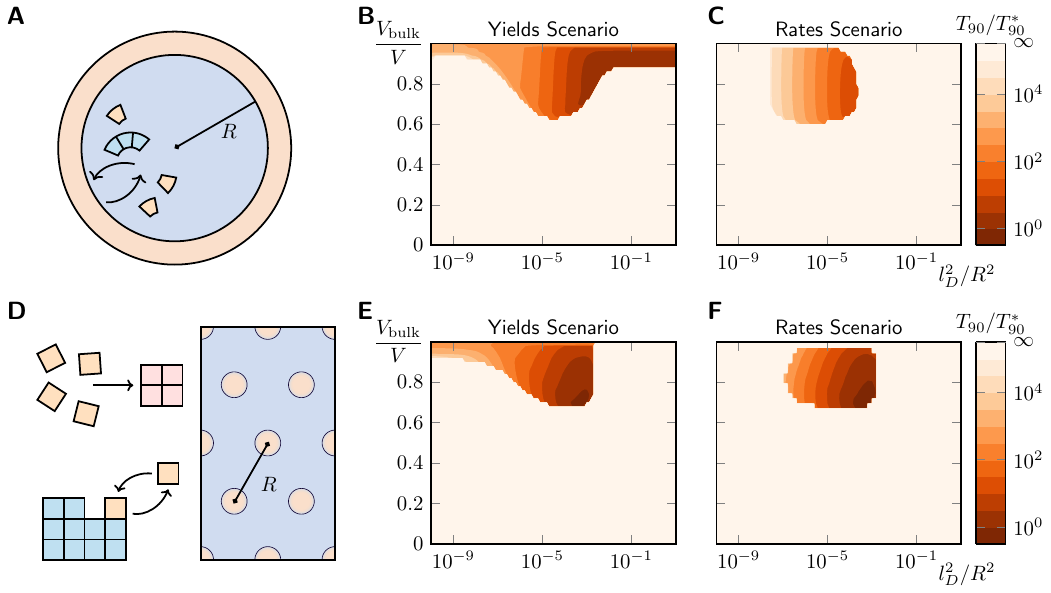}
    \caption{
    \justifying
    Assembly times in two spatially extended systems (analogous to Fig.\@~\ref{fig:results:highlighted-model-extensions}):
    (\textit{A}) A 2D circular domain (bulk; slow compartment) with a 1D boundary (membrane; fast compartment) where the volume fraction $V_S$ is tuned through the radius $R$ and the exchange $\mathcal{D}$ through bulk diffusion $D$ and attachment/detachment rates to/from the membrane.
    (\textit{B}) Corresponding assembly time in the yields scenario (Fig.\@~\ref{fig:results:main-results}\emph{E}).
    (\textit{C}) Corresponding assembly time in the rates scenario (Fig.\@~\ref{fig:results:main-results}\emph{F}).
    (\textit{D}) Different assembly system with square-shaped subunits assembling in a 2D rectangular (slow) bulk domain with (fast) circular domains, e.g., surface patterns, where the volume is set by the total area of the circular domains and exchange by diffusion and the inter-condensate distance.
    (\textit{E}) Corresponding assembly time in the yields scenario (Fig.\@~\ref{fig:results:main-results}\emph{E}).
    (\textit{F}) Corresponding assembly time in the rates scenario (Fig.\@~\ref{fig:results:main-results}\emph{F}).
    The final yield corresponding to (\textit{B}, \textit{C}, \textit{E}, \textit{F}) and are shown in Fig.\@~\ref{fig:results:highlighted-model-extensions}.
    The data for ${V_{\mathrm{bulk}}=0}$ (and ${V_{\mathrm{bulk}}=V}$) shows the corresponding well-mixed system with the parameters of the fast (or slow) domain.
    See \materialsAndMethodsLabel{} for the corresponding sets of equations and details of the numerical simulations.
    }
    \label{si:fig:assembly_times_for_model_extensions}
\end{figure*}

\begin{figure*}[h]
    \centering
    \includegraphics[]{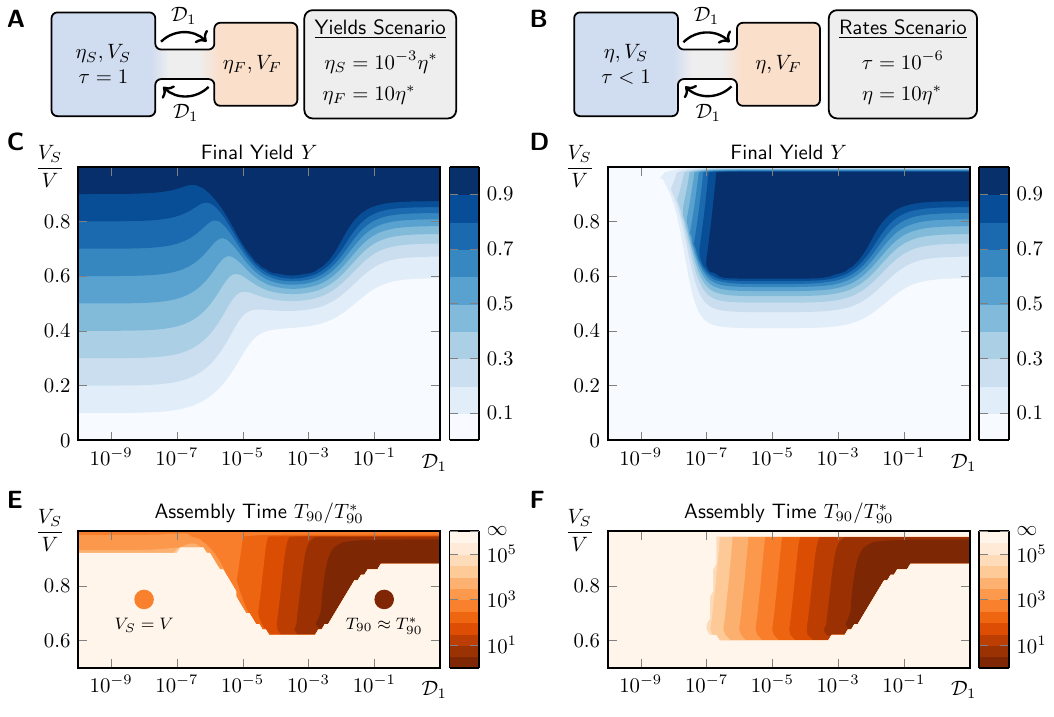}
    \caption{
        \justifying
        Results for the yields and rates scenario in two compartments with only subunits exchanging (analogous to Fig.\@~\ref{fig:results:main-results}):
        (\textit{A}, \textit{B}) Non-dimensionalized control parameters and exemplary parameter choices for the remaining panels for the yields scenario and rates scenario, respectively.
        (\textit{C}) Contour plot of the final yield for varying subunit exchange rates~$\mathcal{D}_1$ and volume ratios $V_S/V$ in the yields scenario.
        (\textit{D}) Same for rates scenario.
        (\textit{E}) Contour plot for the assembly time in the yields scenario normalized with the ideal well mixed assembly time $T_{90}^*$.
        The two inset points are a guide for the eye to compare with the times reached with only the slow (high-yield) compartment~(${V_S=V}$) and the ideal well-mixed time which is realized if ${\bar\eta=\eta^*}$ and ${\mathcal{D}_1=\infty}$.
        (\textit{F}) Same for rates scenario.
        (\textit{C}--\textit{F}) The data for ${V_S=0}$ (${V_S=V}$) corresponds to a single well-mixed system with the parameters of the fast (slow) compartment.
    }
    \label{si:fig:two_compartment_results_only_subunits_exchanging}
\end{figure*}

\begin{figure*}[h]
    \centering
    \includegraphics[]{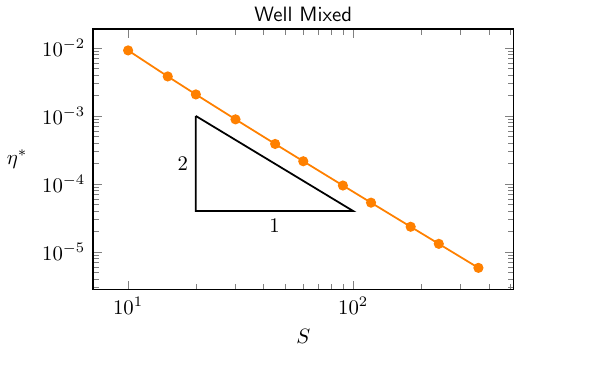}
    \caption{
    \justifying
    Optimal nucleation-to-growth ratio $\eta^*$ as a function of the structure size $S$ in a single well-mixed compartment. The optima approximately scale as $\eta^* \sim S^{-2}$.}
    \label{si:fig:optimal_eta_scales_as_square_of_structure_size}
\end{figure*}

\end{document}